\documentclass[3p,sort&compress]{elsarticle}

\usepackage{amsmath,amssymb}
\usepackage{mathtools}
\usepackage{dsfont}
\usepackage[T1]{fontenc}
\usepackage[utf8]{inputenc}
\usepackage[UKenglish]{babel}
\usepackage{siunitx}
\usepackage{placeins}
\usepackage{bm}
\usepackage{graphicx}
\usepackage[final]{changes}
\graphicspath{{pics/}}
\usepackage{slashed}

\journal{Computer Physics Communication}

\providecommand{\repositoryInformationSetup}{} 
\repositoryInformationSetup

\usepackage{xspace}
\usepackage{bbm}

\newcommand{\Secref}[1]{Section~\ref{sec:#1}}

\newcommand{\figref}[1]{Fig.~\ref{fig:#1}\xspace}
\newcommand{\Figref}[1]{Figure~\ref{fig:#1}\xspace}
\renewcommand{\eqref}[1]{(\ref{eq:#1})\xspace}

\renewcommand{\Ref}[1]{Ref.~\cite{#1}}
\newcommand{\Refs}[1]{Refs.~\cite{#1}}

\newcommand{\goesto}{\ensuremath{\rightarrow}}
\newcommand{\infinity}{\infty}

\newcommand{\order}[1]{\ensuremath{\mathcal{O}\left(#1\right)}\xspace}

\newcommand{\Reals}{\mathbb{R}\xspace}

\newcommand{\oneover}[1]{\ensuremath{\frac{1}{#1}}}                             
\newcommand{\inverse}{\ensuremath{^{-1}}}                                       
\newcommand{\half}{\ensuremath{\frac{1}{2}} }                                   

\newcommand{\average}[1]{\ensuremath{\left\langle #1 \right\rangle}\xspace}

\newcommand{\transpose}{\ensuremath{{}^{\top}}}

\newcommand{\eto}[1]{\ensuremath{\mathrm{e}^{#1}}}
\newcommand{\md}{\ensuremath{\mathrm{d}}}

\newcommand{\del}[2]{\ensuremath{\frac{\partial #1}{\partial#2}}}

\newcommand{\tK}{\ensuremath{\tilde{K}}}

\let\builtinLaTeX\LaTeX
\def\LaTeX{\builtinLaTeX\xspace}
\address[bonn]{
	Helmholtz-Institut f\"{u}r Strahlen- und Kernphysik,
	Rheinische Friedrich-Wilhelms-Universit\"{a}t Bonn, 53012 Bonn Germany
}

\address[fzj]{
	Institut f\"{u}r Kernphysik \& Institute for Advanced Simulation,
	Forschungszentrum J\"{u}lich, 54245 J\"{u}lich Germany
}

\address[mcfp]{
    Maryland Center for Fundamental Physics,
    University of Maryland, College Park 20742, USA
}

\address[cyprus]{
    Computation-based  Science  and  Technology  Research  Center,
    The  Cyprus  Institute,  20  Kavafi  Street,  2121  Nicosia,  Cyprus
}

\begin{document}

\title{The Ising Model with Hybrid Monte Carlo}

\author[bonn]{Johann Ostmeyer}
\author[fzj,mcfp]{Evan Berkowitz}
\author[fzj]{Thomas Luu}
\author[bonn]{Marcus Petschlies}
\author[cyprus]{Ferenc Pittler}

\date{\today}

\begin{abstract}
	The Ising model is a simple statistical model for ferromagnetism.
    There are analytic solutions for low dimensions and very efficient Monte Carlo methods, such as cluster algorithms, for simulating this model in special cases.
    However most approaches do not generalise to arbitrary lattices and couplings.
    We present a formalism that allows one to apply Hybrid Monte Carlo (HMC) simulations to the Ising model, demonstrating how a system with discrete degrees of freedom can be simulated with continuous variables.
    Because of the flexibility of HMC, our formalism is easily generalizable to arbitrary modifications of the model, creating a route to leverage advanced algorithms such as shift preconditioners and multi-level methods, developed in conjunction with HMC.
\end{abstract}

\begin{keyword}
	Ising model \sep HMC \sep Hubbard-Stratonovich transformation \sep lattice Monte Carlo \sep critical slowing down
\end{keyword}

\maketitle

\section{Introduction}\label{sec:intro}

The Ising model is a simple model of ferromagnetism and exhibits a phase transition in dimensions $d\ge 2$.
Analytic solutions determining the critical temperature and magnetization are known for $d=1$ and 2 \cite{onsager_2d_solution}, and in large dimensions the model serves as an exemplary test bed for application of mean-field techniques.
It is also a popular starting point for the discussion of the renormalization group and calculation of critical exponents.

In many cases systems that are seemingly disparate can be  mapped into the Ising model with slight modification.
Examples include certain neural networks~\cite{10.1007/978-3-642-61850-5_18,schneidman:2006}, percolation~\cite{1969PSJJS..26...11K,FORTUIN1972536,Saberi_2010}, ice melt ponds in the arctic~\cite{Ma_2019}, financial markets~\cite{Chowdhury1999AGS,KAIZOJI2002441,SORNETTE2006704}, and population segregation in urban areas\cite{doi:10.1080/0022250X.1971.9989794,doi:10.1119/1.2779882}, to name a few.
In short, the applicability of the Ising model goes well beyond its intended goal of describing ferromagnetic behavior.  Furthermore, it serves as an important pedagogical tool---any serious student of statistical/condensed matter physics as well as field theory should be well versed in the Ising model.

The pedagogical utility of the Ising model extends into numerics as well.
Stochastic lattice methods and the Markov-chain Monte-Carlo (MCMC) concept are routinely introduced via application to the Ising model.
Examples range from the simple Metropolis-Hastings algorithm to more advanced cluster routines, such as Swendsen-Wang~\cite{Swendsen:1987ce} and Wolff~\cite{Wolff:1988uh} and the worm algorithm of Prokof'ev and Svistunov~\cite{PhysRevLett.87.160601}.
Because so much is known of the Ising model, it also serves as a standard test bed for novel algorithms.
Machine learning (ML) techniques were recently applied to the Ising model to aid in identification of phase transitions and order parameters~\cite{Wetzel:2017ooo,Cossu:2018pxj,JMLR:v18:17-527,GIANNETTI2019114639,Alexandrou:2019hgt}.

A common feature of the algorithms mentioned above is that they are well suited for systems with discrete internal spaces, which of course includes the Ising model.
For continuous degrees of freedom the hybrid Monte Carlo (HMC) algorithm~\cite{Duane:1987de} is instead the standard workhorse.
Lattice quantum chromodynamics (LQCD) calculations, for example, rely strongly on HMC.
Certain applications in condensed matter physics now also routinely use HMC~\cite{Luu:2015gpl,Krieg:2018pqh,Wynen:2018ryx}.
Furthermore, algorithms related to preconditioning and multi-level integration have greatly extended the efficacy and utility of HMC.
With the need to sample posterior distributions in so-called big data applications, HMC has become widespread even beyond scientific applications.

It is natural to ask, then, how to apply the numerically-efficient HMC to the broadly-applicable Ising model.
At first glance, the Ising model's discrete variables pose an obstacle for smoothly integrating the Hamiltonian equations of motion to arrive at a new proposal.
However, in Ref.~\cite{pakman2015auxiliaryvariable} a modified version of HMC was introduced where sampling was done over a mixture of binary and continuous distributions and successfully benchmarked to the Ising model in 1D and 2D.  In our work, we describe how to transform the Ising model to a completely continuous space in arbitrary dimensions and with arbitrary couplings between spins (and not just nearest neighbor couplings). \added{Some of these results have already been published in Ref.~\cite{similar_work} without our knowledge and have thus been `rediscovered' by us. Yet, we propose a novel, more efficient approach for the transformation and we perform a thorough analysis of said efficiency and the best choice of the tunable parameter.}

Furthermore, we hope this paper serves a pedagogical function, as a nice platform for introducing both HMC and the Ising model, and a clarifying function, demonstrating how HMC can be leveraged for models with discrete internal spaces.
So, for pedagogical reasons, our implementation of HMC is the simplest `vanilla' version.
As such, it does not compete well, in the numerical sense, with the more advanced cluster algorithms mentioned above.
However, it seems likely that by leveraging the structure of the Ising model one could find a competitive HMC-based algorithm, but we leave such investigations for the future.

This paper is organized as follows.
In \Secref{formalism} we review the Ising model.
We describe how one can transform the Ising model, which resides in a discrete spin space, into a model residing in a continuous space by introducing an auxiliary field and integrating out the spin degrees of freedom.
The numerical stability of such a transformation is not trivial\footnote{Such stability considerations have been egregiously ignored in the past.}, and we describe the conditions for maintaining stability.
With our continuous space defined, we show in \Secref{hmc} how to simulate the system with HMC.
Such a discussion of course includes a cursory description of the HMC algorithm.
In \Secref{results} we show how to calculate observables within this continuous space, since quantities such as magnetization or average energy are originally defined in terms of spin degrees of freedom which are no longer present. We also provide numerical results of key observables, demonstrating proof-of-principle.
We conclude in \Secref{conclusion}.


\section{Formalism}\label{sec:formalism}

The Ising model on a lattice with $N$ sites is described by the Hamiltonian
\begin{align}
    H
    &=
        -   J\sum_{\average{i,j}}s_is_j
        -   \sum_i h_is_i
    \\
    &=
        -   \half J s\transpose K s^{}
        -   h\cdot s
        \label{eq:hamilton_matrix_notation}
\end{align}
where $s_i=\pm1$ are the spins on sites $i = 1,\ldots,N$, $J$ the coupling between neighbouring spins (denoted by $\average{i,j}$),  $h_i$ is the local external magnetic field, and the $\transpose$ superscript denotes the transpose.
We also define the symmetric connectivity matrix $K$ containing the information about the nearest neighbour couplings.
The factor $\half$ on the nearest-neighbor term~\eqref{hamilton_matrix_notation} accounts for the double counting of neighbour pairs that arises from making $K$ symmetric.
If $h$ is constant across all sites we write
\begin{equation}
    \label{eq:constant_h}
    h
    =
    h_0\left(\begin{array}{c}
            1       \\
            1       \\
            \vdots  \\
            1
        \end{array}\right)
        \,.
\end{equation}
We assume a constant coupling $J$ for simplicity in this work.
The same formalism developed here can however be applied for site-dependent couplings as well.
In this case we simply have to replace the matrix $JK$ by the full coupling matrix.

The partition sum over all spin configurations $\left\{ s_i \right\} \equiv \left\{ s_i \,|\, i = 1,\ldots,N \right\}$
\begin{align}
    Z
    =
    \sum_{\left\{s_i\right\}=\pm 1} \eto{-\beta H}
\end{align}
with the inverse temperature $\beta$ is impractical to compute directly for large lattices because the number of terms increases exponentially,  providing the motivation for Monte Carlo methods.
Our goal is to rewrite $Z$ in terms of a continuous variable so that molecular dynamics (MD) becomes applicable.
The usual way to eliminate the discrete degrees of freedom and replace them by continuous ones is via the Hubbard-Stratonovich (HS) transformation.
For a positive definite matrix $A\in\Reals^{N\times N}$ and some vector $v\in\Reals^N$,  the HS relation reads
\begin{align}
    \eto{\half v\transpose A v}
    =
    \frac{1}{\sqrt{\det A\left(2\pi\right)^N}}
    \int\limits_{-\infty}^{\infty}\left[\prod_{i=1}^{N}\md \phi_i\right]\,
        \eto{-\half \phi\transpose A\inverse\phi^{}+v\cdot \phi}
\end{align}
where we integrate over an \emph{auxiliary field} $\phi$.
The argument of the exponent has been linearized in $v$.
In our case the matrix $J'K$ with
\begin{align}
J'\coloneqq\beta J\label{eq:def_J_prime}
\end{align}
takes the place of $A$ in the expression above.
However, $J'K$ is not positive definite in general, nor is $-J'K$.
The eigenvalues $\lambda$ of $K$ are distributed in the interval
\begin{align}
    \lambda
    \in
    \left[-n,\,n\right]\label{eq:eigen_spectrum_K}
\end{align}
where $n$ is the maximal number of nearest neighbours a site can have.
In the thermodynamic limit $N\goesto\infty$ the spectrum becomes continuous and all values in the interval are reached.
Thus the HS transformation is not stable: the Gaussian integral with negative eigenvalues does not converge.

We have to modify the connectivity matrix in such a way that we can apply the HS transformation.
Therefore we introduce a constant shift $C$ to the $K$ matrix,
\begin{align}
    \tilde{K}
    \coloneqq
    K + C\,\mathds{1},
\end{align}
where $C$ has to have the same sign as $J'$, by adding and subtracting the corresponding term in the Hamiltonian.
Now $\tilde{K}$ has the same eigenspectrum as $K$, but shifted by $C\,$.
Thus if we choose $\left|C\right|>n$, $J'\tK$ is positive definite.
We will take such a choice for granted from now on.
For variable coupling the interval~\eqref{eigen_spectrum_K} might have to be adjusted, but the eigenspectrum remains bounded from below, so $C$ can be chosen large enough to make $J'\tK$ positive definite.

Now we can apply the HS transformation to the partition sum
\begin{align}
    Z
    &=
        \sum_{\left\{s_i\right\}=\pm 1}
            \eto{
                    \half\beta Js\transpose \tK s{}
                -   \half \beta J C s^2
                + \beta h\cdot s
                }
    \\
    &=
        \eto{-\half J' C N}
        \sum_{\left\{s_i\right\}=\pm 1}
            \eto{
                    \half J's\transpose \tK s^{}
                +   h'\cdot s
                }
    \label{eq:absorb_beta}\\
    &=
        \eto{-\half J' C N}
        \sum_{\left\{s_i\right\}=\pm 1}
            \frac{1}{\sqrt{\det \tK\left(2\pi J'\right)^N}}
            \int\limits_{-\infty}^{\infty}\left[\prod_{i=1}^{N}\md \phi_i\right]\,
                \eto{
                    -   \oneover{2J'}\phi\transpose\tK\inverse \phi^{}
                    +   \left(h'+\phi\right)\cdot s
                    }
    \label{eq:did_HS_trafo}\\
    &=
        \frac{\eto{-\half J' C N}}{\sqrt{\det \tK\left(2\pi J'\right)^N}}
        \int\limits_{-\infty}^{\infty}\left[\prod_{i=1}^{N}\md \phi_i\right]\,
            \eto{ - \oneover{2J'}\phi\transpose\tK\inverse\phi}
            \left[\prod_{i=1}^N 2\cosh\left(h'_i+\phi_i\right)\right]
    \label{eq:summed_over_si}\\
    &=
        2^N
        \frac{\eto{-\half J' C N}}{\sqrt{\det \tK\left(2\pi J'\right)^N}}
        \int\limits_{-\infty}^{\infty}\left[\prod_{i=1}^{N}\md \phi_i\right]\,
            \eto{
                -   \oneover{2J'}\phi\transpose \tK\inverse \phi^{}
                +   \sum_i\log\cosh\left(h'_i+\phi_i\right)
                }
    \label{eq:cosh_in_exponent}
\end{align}
where we used in \eqref{absorb_beta} that $s_i^2=1$ for all $i$ and defined $h'\coloneqq \beta h$ in analogy with~\eqref{def_J_prime}.
In \eqref{did_HS_trafo} we performed the HS transformation and in \eqref{summed_over_si} we explicitly evaluated the sum over all the now-independent $s_i$, thereby integrating out the spins.
After rewriting the $\cosh$ term in \eqref{cosh_in_exponent} we are left with an effective action that can be used to perform HMC calculations.
However, we do not recommend using this form directly, as it needs a matrix inversion.

Instead, let us perform the substitution
\begin{align}
    \phi
    &=
    \sqrt{J'}\tK\psi-h'
\end{align}
with the functional determinant $\sqrt{J'}^N\det\tK$.
This \added{substitution is going to bring a significant speed up and has not been considered in Ref.~\cite{similar_work}. It} allows us to get rid of the inverse of $\tK$ in the variable part of the partition sum
\begin{align}
    Z
    &=
    \sqrt{\left(\tfrac 2\pi\right)^N\det\tK}\:
    \eto{-\half J' C N}
    \int\limits_{-\infty}^{\infty}\left[\prod_{i=1}^{N}\md \psi_i\right]\,
        \eto{
            -   \half\psi\transpose \tK\psi^{}
            +   \oneover{\sqrt{J'}}\psi\cdot h'
            -   \oneover{2J'}h'\transpose \tK\inverse h'^{}
            +   \sum_i\log\cosh\left(
                    \sqrt{J'}\left(\tK\psi\right)_i
                    \right)
            }\,.
    \label{eq:phi_psi_substitution}
\end{align}
The only left over term involving an inversion remains in the constant $h'\transpose\tK\inverse h'^{}$.
Fortunately this does not need to be calculated during HMC simulations.
We do need it, however, for the calculation of some observables, such as the magnetisation \eqref{formula_for_m}, and for this purpose it can be calculated once without any need for updates. \added{Let us also remark that the inverse $\tK^{-1}$ does not have to be calculated exactly. Instead it suffices to solve the system of linear equations $\tK x = h'$ for $x$ which can be done very efficiently with iterative solvers, such as the conjugate gradient (CG) method~\cite{templates}.}

A further simplification can be achieved when the magnetic field is constant \eqref{constant_h} and every lattice site has the same number of nearest neighbours $n_0$.
Then we find that
\begin{align}
    \tK h'
    =
    \left(n_0+C\right)h'
\end{align}
and thus
\begin{equation}
    h'\transpose \tK\inverse h'^{}
    =
    h'\transpose\oneover{n_0+C}h'^{}
    =
    \frac{N}{n_0+C}h_0'^2\,.\label{eq:hKh_for_constant_h}
\end{equation}

\section{HMC}\label{sec:hmc}

Hybrid Monte Carlo\footnote{Sometimes `Hamiltonian Monte Carlo', especially in settings other than lattice quantum field theory.} (HMC)~\cite{Duane:1987de} requires introducing a fictitious \emph{molecular dynamics time} and conjugate momenta, integrating current field configurations according to Hamiltonian equations of motion to make a Metropolis proposal.
We multiply the partition sum $Z$ \eqref{phi_psi_substitution} by unity, using the Gaussian identity
\begin{equation}
	\label{eq:momentum distribution}
	\frac{1}{(2\pi)^{N/2}} \int_{-\infty}^{+\infty} \left[\prod_{i=1}^N \md p_i\right]\, \eto{-\frac{1}{2}p_i^2} = 1
\end{equation}
where we have one conjugate momentum $p$ for each field variable $\psi$ in $Z$, and we sample configurations of fields and momenta from this combined distribution.
The conceptual advantage of introducing these momenta is that we can evolve the auxiliary fields $\psi$ with the HMC Hamiltonian~$\mathcal{H}$,
\begin{align}
    \mathcal{H}
    =
        \half p^2
    +   \half\psi\transpose\tK\psi^{}
    -   \oneover{\sqrt{J'}}\psi\cdot h'
    -   \sum_i\log\cosh\left(\sqrt{J'}\left(\tK\psi\right)_i\right)
\end{align}
by integrating the equations of motion (EOM)
\begin{align}
    \dot{\psi}
    &=  +\del{\mathcal{H}}{p}
     =  p
    \\
    \dot{p}
    &=  -\del{\mathcal{H}}{\psi}
     =  -\tK\psi + \oneover{\sqrt{J'}} h' + \sqrt{J'}\tK\tanh\left(\sqrt{J'}\tK\psi\right)
\end{align}
where the $\tanh$ is understood element-wise.

Thus one can employ the Hybrid Monte Carlo algorithm to generate an ensemble of field configurations by a Markov chain.
Starting with some initial configuration $\psi$, the momentum $p$ is sampled according to a Gaussian distribution \eqref{momentum distribution}.
The EOM are integrated to update all the field variables at once.
The integration of the differential equations, or the \emph{molecular dynamics}, is performed by a (volume-preserving) symmetric symplectic integrator (we use leap-frog here, but more efficient schemes can be applied \cite{PhysRevE.65.056706,OMELYAN2003272}) to ensure an unbiased update.
The equations of motion are integrated one \emph{molecular dynamics time unit}, which is held fixed for each ensemble, to produce one \emph{trajectory} through the configuration space; the end of the trajectory is proposed as the next step in the Markov chain.
If the molecular dynamics time unit is very short, the new proposal will be very correlated with the current configuration. If the molecular dynamics time unit is too long, it will be very expensive to perform an update.

The proposal is accepted with the Boltzmann probability $\min\left(1,\,\eto{-\Delta \mathcal{H}}\right)$ where the energy splitting $\Delta \mathcal{H}=\mathcal{H}_\text{new}-\mathcal{H}_\text{old}$ is the energy difference between the proposed configuration and the current configuration.
If our integration algorithm were exact, $\Delta \mathcal{H}$ would vanish and we would always accept the new proposal, by conservation of energy.
The Metropolis-Hastings accept/reject step guarantees that we get the correct distribution despite inexact numerical integration.
So, if we integrate with time steps that are too coarse we will reject more often.  Finer integration ensures a greater acceptance rate, all else being equal.

If the proposal is not accepted as the next step of our Markov chain, it is rejected and the previous configuration repeats.
After each accepted or rejected proposal the momenta are refreshed according to the Gaussian distribution \eqref{momentum distribution} and molecular dynamics integration resumes, to produce the next proposal.

If the very first configuration is not a good representative of configurations with large weight, the Markov chain will need to be \emph{thermalized}---driven towards a representative place---by running the algorithm for some number of updates.
Then, production begins.
An ensemble of $N_\text{cf}$ configurations $\left\{\psi^n\right\}$ is drawn from the Markov chain and the estimator of any observable $O(\psi)$
\begin{align}
    \overline{O}
    &=
    \oneover{N_\text{cf}}\sum\limits_{n=1}^{N_\text{cf}}O\!\left(\psi^n\right)
    \\
    &\text{converges to the expectation value} \nonumber
    \\
    \average{O}
    &= \oneover{Z} \sqrt{\left(\tfrac 2\pi\right)^N\det\tK}\:
    \eto{-\half J' C N}
    \int\limits_{-\infty}^{\infty}\left[\prod_{i=1}^{N}\md \psi_i\right]\,
        O(\psi)\;
        \eto{
            -   \half\psi\transpose\tK\psi^{}
            +   \oneover{\sqrt{J'}}\psi\cdot h'
            -   \oneover{2J'}h'\transpose\tK\inverse h'^{}
            +   \sum_i\log\cosh\left(
                    \sqrt{J'}\left(\tK\psi\right)_i
                    \right)
            }
\end{align}
as the ensemble size $N_\text{cf}\goesto\infinity$, with uncertainties on the scale of $N_\text{cf}^{-1/2}$ as long as the configurations are not noticeably correlated---if their \emph{autocorrelation time} (in Markov chain steps) is short enough.

Not much time has been spent on the tuning of $C$ during this work.
We expect that the choice of $C$ can influence the speed of the simulations.
Clearly $\left|C\right|$ must not be chosen too large because in the limit $\left|C\right|\goesto\infty$ the Hamiltonian can be approximated by
\begin{align}
    \oneover{C}\mathcal{H}
    &=
            \half \psi^2
        -   \sqrt{J'}\sum_i\left|\psi_i\right|
        +   \order{C^{-1}}
\end{align}
with the minima
\begin{align}
	\label{eq:large C minima}
    \psi_i
    &=
    \pm\sqrt{J'}.
\end{align}
Any deviation from a minimum is enhanced by the factor of $C$ and is thus frozen out for large $\left|C\right|$.
This reproduces the original discrete Ising model up to normalisation factors.
Plainly the HMC breaks down in this case.
As the limit is approached, the values for the $\psi_i$ become confined to smaller and smaller regions.
The result is that HMC simulations can get stuck in local minima and the time series is no longer \emph{ergodic}---it cannot explore all the states of the Markov chain---which may yield incorrect or biased results.
From now on we use $\left|C\right|=n+\num{e-5}$; we later show the effect of changing $C$ in \Figref{compare_C}.

A large coupling (or low temperature) $J'$ introduces an ergodicity problem as well: as we expect to be in a magnetized phase, all the spins should be aligned and flipping even one spin is energetically disfavored even while flipping them all may again yield a likely configuration.
This case however is less problematic because there are only two regions with a domain wall between them; the region with all $\psi_i>0$ and the region with all $\psi_i<0$.
The ergodicity issue is alleviated by proposing a global sign flip and performing a Metropolis accept/reject step every few trajectories, similar to that proposed in Ref.~\cite{Wynen:2018ryx}.

\section{Results}\label{sec:results}

Let us again assume constant external field with strength $h_0$ \eqref{constant_h}.
Then the expectation value of the average magnetisation and energy per site read
\begin{align}
    \average{m}
    &   =   \oneover{NZ}\del{Z}{h'}
    \\
    &   =   \oneover{N}\average{\oneover{\sqrt{J'}}\sum_i\psi_i
        -   \frac{N}{n_0+C}\frac{h'_0}{J'}}
    \\
    &   =   \frac{\average{\psi}}{\sqrt{J'}}
        - \oneover{n_0+C}\frac{h'_0}{J'}\,,\label{eq:formula_for_m}\\
    \average{\beta\varepsilon}
    &	=	-\frac{\beta}{NZ}\del{Z}{\beta}\label{eq:ee form}\\
    &	=	\half CJ' + \oneover{n_0+C}\frac{{h'_0}^2}{2J'} - \frac{h'_0}{2\sqrt{J'}}\average{\psi} - \frac{\sqrt{J'}}{2N}\average{\left(\tK \psi\right)\cdot\tanh\left(\sqrt{J'}\tK\psi\right)}\label{eq:formula_for_e}
\end{align}
where $\average{\psi} = \average{\psi_i}$ for any site $i$ due to translation invariance.
Any other physical observables can be derived in the same way.
For example, higher-point correlation functions like spin-spin correlators may be derived by functionally differentiating with respect to a site-dependent $h_i$ (without the simplification of constant external field \eqref{hKh_for_constant_h}). 
We stress here that, although $C$ appears in observables (as in the magnetization~\eqref{formula_for_m} and energy density~\eqref{formula_for_e}), the results are independent of $C$---its value only influences the convergence rate.

\begin{figure}[ht]
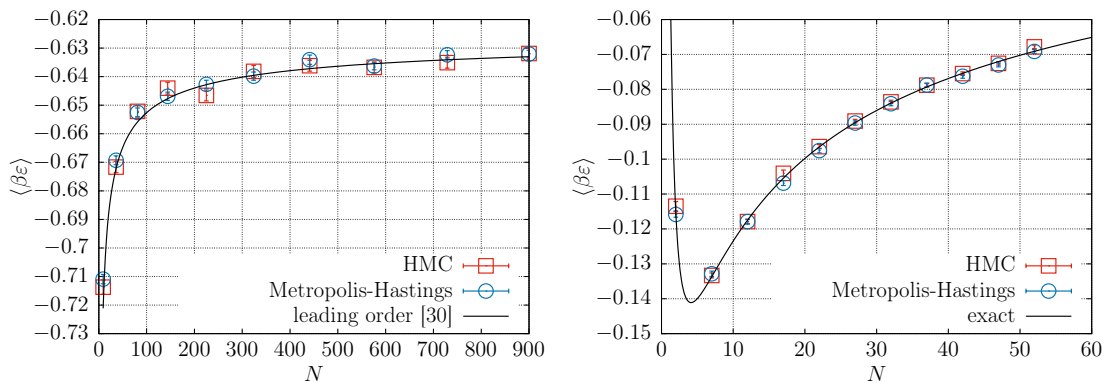

	\centering
	\resizebox{0.443\textwidth}{!}{{\Large\input{pics/energy_2d}}}
	\resizebox{0.443\textwidth}{!}{{\Large\input{pics/energy_alltoall}}}
	\caption{Expectation value of the energy per site for the two dimensional periodic square lattice (left) and the lattice with all-to-all coupling (right) for the HMC and the Metropolis-Hastings algorithms at critical coupling and $h=0$ with lattice sizes $N$.}
	\label{fig:magnetisation}
\end{figure}

In \Figref{magnetisation} we demonstrate that the HMC algorithm\footnote{Our code is publicly available under \url{https://github.com/HISKP-LQCD/ising_hmc}.} indeed produces correct results.
The left panel shows the average energy per site at the critical point~\cite{onsager_2d_solution} of the two-dimensional square lattice with periodic boundary conditions.
We choose to scale the number of integration steps per trajectory with the lattice volume as 
$N_\text{step} = \left\lfloor\log N\right\rfloor$, which empirically leads to acceptance rates between 70\% and 80\% for a broad range of lattice sizes and dimensions.
The results from the HMC simulations are compared to the results obtained via the local Metropolis-Hastings algorithm with the same number $N_\text{cf}$ of sweeps (a sweep consists of $N$ spin flip proposals).
In addition we show the leading order analytic results \cite{ising_2d_exact} $\average{\beta\varepsilon}\approx-\log\left(1+\sqrt{2}\right)\left(\frac{\sqrt{2}}{2}+\frac{1}{3\sqrt{N}}\right)+\order{N^{-1}}$.
We not only find that the results are compatible, but also that the errors of both stochastic methods are comparable.  The right panel shows the average energy per site in the case where the coupling is no longer nearest neighbor, but the extreme opposite with all-to-all couplings.  The Hamiltonian we use in this case is, up to an overall constant, the ``infinite-range'' Ising model~\cite{negele1988quantum}.  This model has analytic solutions for physical observables as a function of the number of lattice sites $N$ which we show for the case of the average energy (black line).  We provide a description of this model, as a well as a derivation of the exact solution for the average energy, in \ref{sect:infiniteRangeIsing}.  Our numerical results agree very well with the exact result.

\begin{figure}[ht]
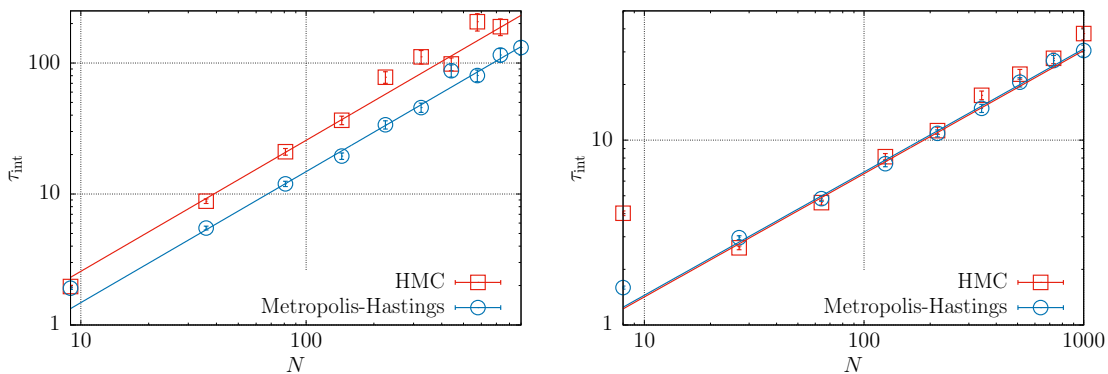

	\centering
	\resizebox{0.443\textwidth}{!}{{\Large\input{pics/autocorrelation}}}
	\resizebox{0.443\textwidth}{!}{{\Large\input{pics/autocorrelation_3d}}}
	\caption{Integrated autocorrelation time of $|m|$ for the HMC and the Metropolis-Hastings algorithms at critical coupling and $h=0$ for the $d=2$ (left) and $d=3$ (right) dimensional periodic square lattice with size $N$. The lines are fits of the form $\tau_\text{int}=\alpha N^\frac 2d$ for $N>10$.}
	\label{fig:tau_int}
\end{figure}

Since it is not the aim of this work to present physical results, but rather to introduce an alternative formulation for simulating the Ising model and generalizations thereof,
we do not compute other observables explicitly, nor do we investigate their dependence on other parameters.
On the other hand it is not sufficient that the algorithm in principle produces correct results---we must also investigate its efficiency.
A good measure for the efficiency is the severity of \emph{critical slowing down}---that the integrated autocorrelation time\footnote{$\tau_\text{int}$ and its error have been calculated according to the scheme proposed in Ref~\cite{monte_carlo_errors}.} $\tau_\text{int}$ diverges at the critical point as some power $\gamma$ of the system size $\tau_\text{int}\propto N^\gamma$.
One could expect that, being a global update algorithm, the HMC does not suffer as much from critical slowing down as Metropolis-Hastings.
\Figref{tau_int} however shows that both algorithms have dynamic exponent $z \equiv d\gamma \approx 2$ in $d=2$ and $d=3$ dimensions (see Ref.~\cite{Pelissetto:2000ek} and references within for a discussion of the critical coupling and exponents in $d=3$).
Still one has to keep in mind that a Metropolis-Hastings sweep takes less time than an HMC trajectory and the HMC trajectories become logarithmically longer as $N$ grows.
In our implementation we find the proportionality
	\begin{align}
		T_\text{HMC}\approx 4N_\text{step}T_\text{MH}
	\end{align}
where $T_\text{HMC}$ is the time required for one HMC trajectory and $T_\text{MH}$ the time required for one Metropolis-Hastings sweep.

\begin{figure}[ht]
		\centering
		\resizebox{0.443\textwidth}{!}{{\Large\input{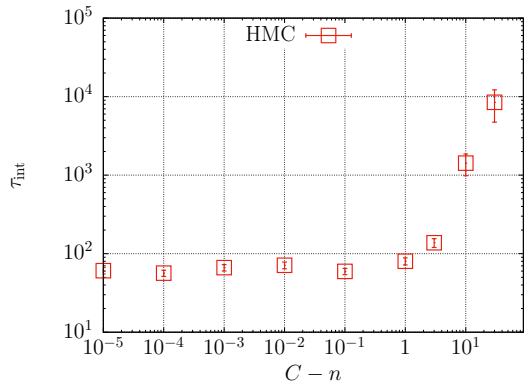}}}
		\caption{Integrated autocorrelation time of $|m|$ for the HMC algorithm at critical coupling and $h=0$ for the $d=2$ dimensional periodic square lattice with size $N=15^2$ against the shift $C$ reduced by the number of nearest neighbours $n$.}
		\label{fig:compare_C}
	\end{figure}

Last but not least let us study the impact of the shift parameter $C$ on the efficiency of the algorithm by means of the 
  autocorrelation for the absolute magnetization $|m|$.
As explained earlier, when $C$ becomes very large the potential becomes very steep around the local minima \eqref{large C minima}.
When this localization becomes important we expect the autocorrelation to increase with $C$, as transitions from one local minimum to another become less likely.
This behaviour can be seen in \Figref{compare_C}.
We find that the autocorrelation is constant within errors below some critical value, in this case $C_\text{crit}\approx n+1$, and increases rapidly for larger $C$.
So, as long as the potential is not too deep HMC can explore the whole configuration space.
A very large $C$ causes wells from which it is difficult to escape, while $C$s just large enough to ensure stability yield very flat, smooth potentials.
We see in \figref{compare_C} that as long as the shift is small $|C|-n \ll 1$ its specific value is irrelevant and does not need to be tuned.
	
\section{Conclusion}\label{sec:conclusion}

In this paper we showed how to apply the HMC algorithm to the Ising model, successfully applying an algorithm that uses only continuous state variables to a system with discrete degrees of freedom.
We find that the HMC algorithm generalises the Ising model very well to arbitrary geometries without much effort.
It has been presented here in the most simple form.
In this simple form the HMC is an extremely inefficient algorithm if applied to the Ising model.
Although more flexible than the most efficient methods, such as cluster algorithms, it loses as compared even to the Metropolis-Hastings algorithm.
The coefficient by which the Metropolis-Hastings algorithm surpasses the HMC decreases with dimension, so that HMC might be preferable in case of an extremely high number of nearest neighbours---in the case of less local coupling, for example.

Moreover, for physical systems that suffer from sign problems, one may hope to leverage complex Langevin, Lefschetz thimble, or other contour-optimizing methods (for a dramatically incomplete set of examples, consider, respectively, \Ref{Aarts:2010gr,Fodor:2015doa}, \Refs{Cristoforetti:2012su,Fujii:2013sra,Alexandru:2017oyw}, and \Refs{Alexandru:2017czx,Mori:2017pne,Kashiwa:2018vxr} and references therein).
The formulation in terms of continuous variables presented here is well-suited for these methods, while the methods that deal directly with the original discrete variables such as the Metropolis-Hastings, cluster, and worm algorithms, for example, are non-starters.
In that sense, our exact reformulation and HMC method can be seen as the first step towards solving otherwise-intractable problems.

The HMC algorithm could be optimised by more efficient integrators and different choices of $C$, just to name the most obvious possibilities.
Many more methods have been developed to improve HMC performance and it is expected that some of them could also speed up the Ising model.


\section*{Acknowledgements}

We thank Paulo Bedaque, Matthias Fischer, Michael Kajan, Ulf Mei{\ss}ner, Marcel Nitsch, Carsten Urbach and Jan-Lukas Wynen for their constructive criticisms.  This work was done in part through financial support from the Deutsche Forschungsgemeinschaft (Sino-German CRC 110).
E.B. is supported by the U.S. Department of Energy under Contract No. DE-FG02-93ER-40762.

\appendix

\section{The ``infinite-range Ising model''}\label{sect:infiniteRangeIsing}

For the calculations shown on the right panel of Figure~\ref{fig:magnetisation} we used the following Hamiltonian,
\begin{align}
H({\bm s})&=-\frac{1}{2}\frac{J}{N}\sum_{i\ne j} s_is_j-h\sum_is_i\\
&=\frac{1}{2}J-\frac{1}{2}\frac{J}{N}\sum_{i, j} s_is_j-h\sum_is_i\ ,
\end{align}
where in the second line we used the fact that $s_i^2=1\ \forall i$ and there is no restriction in the sum over spin couplings.  With the exception of the \emph{self-energy} term $\frac{1}{2}J$ in the second line above,  the remaining terms  constitute the ``infinite-range Ising model''\cite{negele1988quantum}.  From now on we assume that $J>0$, but a similar calculation can be done for $J<0$.  

The partition function for this Hamiltonian can be exactly determined.  Applying the HS transformation as described in the equations leading up to the partition sum \eqref{cosh_in_exponent} one obtains
\begin{equation}
Z=e^{\frac{1}{2}\beta J}\int_{-\infty}^{\infty} \frac{d \phi}{\sqrt{2 \pi \beta \hat{J}}} e^{-\frac{\phi^{2}}{2 \beta \hat{J}}}[2 \cosh (\beta h \pm \phi)]^{N}\ ,
\end{equation}
where we define $\hat J=J/N$.  Expanding the $\cosh$ terms into exponentials allows one to formally integrate over the HS fields, obtaining
\begin{equation}\label{eq:all to all partition function}
Z=e^{\frac{1}{2}\beta J}\sum_{n=0}^{N}\begin{pmatrix}
N \\
n
\end{pmatrix} 
f(\beta \hat{J}, \beta h, N-2 n)\ ,
\end{equation}
and
\begin{equation}
f(\beta \hat{J}, \beta h, x) \equiv e^{\frac{1}{2} \beta \hat{J} x^{2}+\beta h x}\ .
\end{equation}
Using the definition of the energy density \eqref{ee form} with the partition sum \eqref{all to all partition function} gives our analytic expression for the internal energy,
\begin{equation}
\langle\beta \varepsilon\rangle=\frac{1}{2}\beta\hat J-\frac{\beta}{N Z}
\sum_{n=0}^{N}
\begin{pmatrix}
N \\
n
\end{pmatrix}\left[\frac{1}{2}\hat{J}(N-2n)^2+h(N-2n)\right]f(\beta \hat{J}, \beta h, N-2 n)\ .
\end{equation}
Note the relative sign difference between the terms on the right hand side above.  For sufficiently small $N$ the self-energy term wins out, otherwise the second term dominates.

An analogous calculation for the magnetisation with an extrapolation of $N\goesto \infty$ leads to the critical point $\beta J=1$. We used this value in Figure~\ref{fig:magnetisation}.


\FloatBarrier
\bibliographystyle{elsarticle-num}
\bibliography{master}

\begin{thebibliography}{10}
\expandafter\ifx\csname url\endcsname\relax
  \def\url#1{\texttt{#1}}\fi
\expandafter\ifx\csname urlprefix\endcsname\relax\def\urlprefix{URL }\fi
\expandafter\ifx\csname href\endcsname\relax
  \def\href#1#2{#2} \def\path#1{#1}\fi

\bibitem{onsager_2d_solution}
L.~Onsager, \href{https://link.aps.org/doi/10.1103/PhysRev.65.117}{{Crystal
  Statistics. I. A Two-Dimensional Model with an Order-Disorder Transition}},
  Phys. Rev. 65 (1944) 117--149.
\newblock \href {https://doi.org/10.1103/PhysRev.65.117}
  {\path{doi:10.1103/PhysRev.65.117}}.
\newline\urlprefix\url{https://link.aps.org/doi/10.1103/PhysRev.65.117}

\bibitem{10.1007/978-3-642-61850-5_18}
W.~A. Little, An ising model of a neural network, in: W.~J{\"a}ger, H.~Rost,
  P.~Tautu (Eds.), Biological Growth and Spread, Springer Berlin Heidelberg,
  Berlin, Heidelberg, 1980, pp. 173--179.

\bibitem{schneidman:2006}
E.~Schneidman, M.~Berry, R.~S. et~al., Weak pairwise correlations imply
  strongly correlated network states in a neural population, Nature 440 (2006)
  1007--1012.
\newblock \href {https://doi.org/10.1038/nature04701}
  {\path{doi:10.1038/nature04701}}.

\bibitem{1969PSJJS..26...11K}
P.~W. {Kasteleyn}, C.~M. {Fortuin}, {Phase Transitions in Lattice Systems with
  Random Local Properties}, Physical Society of Japan Journal Supplement 26
  (1969) 11.

\bibitem{FORTUIN1972536}
C.~Fortuin, P.~Kasteleyn,
  \href{http://www.sciencedirect.com/science/article/pii/0031891472900456}{On
  the random-cluster model: I. introduction and relation to other models},
  Physica 57~(4) (1972) 536 -- 564.
\newblock \href {https://doi.org/https://doi.org/10.1016/0031-8914(72)90045-6}
  {\path{doi:https://doi.org/10.1016/0031-8914(72)90045-6}}.
\newline\urlprefix\url{http://www.sciencedirect.com/science/article/pii/0031891472900456}

\bibitem{Saberi_2010}
A.~A. Saberi, H.~Dashti-Naserabadi,
  \href{https://doi.org/10.1209%2F0295-5075%2F92%2F67005}{Three-dimensional
  ising model, percolation theory and conformal invariance}, {EPL} (Europhysics
  Letters) 92~(6) (2010) 67005.
\newblock \href {https://doi.org/10.1209/0295-5075/92/67005}
  {\path{doi:10.1209/0295-5075/92/67005}}.
\newline\urlprefix\url{https://doi.org/10.1209%2F0295-5075%2F92%2F67005}

\bibitem{Ma_2019}
Y.-P. Ma, I.~Sudakov, C.~Strong, K.~M. Golden,
  \href{https://doi.org/10.1088%2F1367-2630%2Fab26db}{Ising model for melt
  ponds on arctic sea ice}, New Journal of Physics 21~(6) (2019) 063029.
\newblock \href {https://doi.org/10.1088/1367-2630/ab26db}
  {\path{doi:10.1088/1367-2630/ab26db}}.
\newline\urlprefix\url{https://doi.org/10.1088%2F1367-2630%2Fab26db}

\bibitem{Chowdhury1999AGS}
D.~Chowdhury, D.~Stauffer, A generalized spin model of financial markets, The
  European Physical Journal B - Condensed Matter and Complex Systems 8 (1999)
  477--482.

\bibitem{KAIZOJI2002441}
T.~Kaizoji, S.~Bornholdt, Y.~Fujiwara,
  \href{http://www.sciencedirect.com/science/article/pii/S0378437102012165}{Dynamics
  of price and trading volume in a spin model of stock markets with
  heterogeneous agents}, Physica A: Statistical Mechanics and its Applications
  316~(1) (2002) 441 -- 452.
\newblock \href {https://doi.org/https://doi.org/10.1016/S0378-4371(02)01216-5}
  {\path{doi:https://doi.org/10.1016/S0378-4371(02)01216-5}}.
\newline\urlprefix\url{http://www.sciencedirect.com/science/article/pii/S0378437102012165}

\bibitem{SORNETTE2006704}
D.~Sornette, W.-X. Zhou,
  \href{http://www.sciencedirect.com/science/article/pii/S0378437106002354}{Importance
  of positive feedbacks and overconfidence in a self-fulfilling ising model of
  financial markets}, Physica A: Statistical Mechanics and its Applications
  370~(2) (2006) 704 -- 726.
\newblock \href {https://doi.org/https://doi.org/10.1016/j.physa.2006.02.022}
  {\path{doi:https://doi.org/10.1016/j.physa.2006.02.022}}.
\newline\urlprefix\url{http://www.sciencedirect.com/science/article/pii/S0378437106002354}

\bibitem{doi:10.1080/0022250X.1971.9989794}
T.~C. Schelling, \href{https://doi.org/10.1080/0022250X.1971.9989794}{Dynamic
  models of segregation}, The Journal of Mathematical Sociology 1~(2) (1971)
  143--186.
\newblock \href
  {http://arxiv.org/abs/https://doi.org/10.1080/0022250X.1971.9989794}
  {\path{arXiv:https://doi.org/10.1080/0022250X.1971.9989794}}, \href
  {https://doi.org/10.1080/0022250X.1971.9989794}
  {\path{doi:10.1080/0022250X.1971.9989794}}.
\newline\urlprefix\url{https://doi.org/10.1080/0022250X.1971.9989794}

\bibitem{doi:10.1119/1.2779882}
D.~Stauffer, \href{https://doi.org/10.1119/1.2779882}{Social applications of
  two-dimensional ising models}, American Journal of Physics 76~(4) (2008)
  470--473.
\newblock \href {http://arxiv.org/abs/https://doi.org/10.1119/1.2779882}
  {\path{arXiv:https://doi.org/10.1119/1.2779882}}, \href
  {https://doi.org/10.1119/1.2779882} {\path{doi:10.1119/1.2779882}}.
\newline\urlprefix\url{https://doi.org/10.1119/1.2779882}

\bibitem{Swendsen:1987ce}
R.~H. Swendsen, J.-S. Wang, {Nonuniversal critical dynamics in Monte Carlo
  simulations}, Phys. Rev. Lett. 58 (1987) 86--88.
\newblock \href {https://doi.org/10.1103/PhysRevLett.58.86}
  {\path{doi:10.1103/PhysRevLett.58.86}}.

\bibitem{Wolff:1988uh}
U.~Wolff, {Collective Monte Carlo Updating for Spin Systems}, Phys. Rev. Lett.
  62 (1989) 361.
\newblock \href {https://doi.org/10.1103/PhysRevLett.62.361}
  {\path{doi:10.1103/PhysRevLett.62.361}}.

\bibitem{PhysRevLett.87.160601}
N.~Prokof'ev, B.~Svistunov,
  \href{https://link.aps.org/doi/10.1103/PhysRevLett.87.160601}{Worm algorithms
  for classical statistical models}, Phys. Rev. Lett. 87 (2001) 160601.
\newblock \href {https://doi.org/10.1103/PhysRevLett.87.160601}
  {\path{doi:10.1103/PhysRevLett.87.160601}}.
\newline\urlprefix\url{https://link.aps.org/doi/10.1103/PhysRevLett.87.160601}

\bibitem{Wetzel:2017ooo}
S.~J. Wetzel, M.~Scherzer, {Machine Learning of Explicit Order Parameters: From
  the Ising Model to SU(2) Lattice Gauge Theory}, Phys. Rev. B96~(18) (2017)
  184410.
\newblock \href {http://arxiv.org/abs/1705.05582} {\path{arXiv:1705.05582}},
  \href {https://doi.org/10.1103/PhysRevB.96.184410}
  {\path{doi:10.1103/PhysRevB.96.184410}}.

\bibitem{Cossu:2018pxj}
G.~Cossu, L.~Del~Debbio, T.~Giani, A.~Khamseh, M.~Wilson, {Machine learning
  determination of dynamical parameters: The Ising model case}, Phys. Rev.
  B100~(6) (2019) 064304.
\newblock \href {http://arxiv.org/abs/1810.11503} {\path{arXiv:1810.11503}},
  \href {https://doi.org/10.1103/PhysRevB.100.064304}
  {\path{doi:10.1103/PhysRevB.100.064304}}.

\bibitem{JMLR:v18:17-527}
A.~Morningstar, R.~G. Melko, \href{http://jmlr.org/papers/v18/17-527.html}{Deep
  learning the ising model near criticality}, Journal of Machine Learning
  Research 18~(163) (2018) 1--17.
\newline\urlprefix\url{http://jmlr.org/papers/v18/17-527.html}

\bibitem{GIANNETTI2019114639}
C.~Giannetti, B.~Lucini, D.~Vadacchino,
  \href{http://www.sciencedirect.com/science/article/pii/S0550321319301257}{Machine
  learning as a universal tool for quantitative investigations of phase
  transitions}, Nuclear Physics B 944 (2019) 114639.
\newblock \href
  {https://doi.org/https://doi.org/10.1016/j.nuclphysb.2019.114639}
  {\path{doi:https://doi.org/10.1016/j.nuclphysb.2019.114639}}.
\newline\urlprefix\url{http://www.sciencedirect.com/science/article/pii/S0550321319301257}

\bibitem{Alexandrou:2019hgt}
C.~Alexandrou, A.~Athenodorou, C.~Chrysostomou, S.~Paul, {Unsupervised
  identification of the phase transition on the 2D-Ising model} (2019).
\newblock \href {http://arxiv.org/abs/1903.03506} {\path{arXiv:1903.03506}}.

\bibitem{Duane:1987de}
S.~Duane, A.~D. Kennedy, B.~J. Pendleton, D.~Roweth, {Hybrid Monte Carlo},
  Phys. Lett. B195 (1987) 216--222.
\newblock \href {https://doi.org/10.1016/0370-2693(87)91197-X}
  {\path{doi:10.1016/0370-2693(87)91197-X}}.

\bibitem{Luu:2015gpl}
T.~Luu, T.~A. Lähde, {Quantum Monte Carlo Calculations for Carbon Nanotubes},
  Phys. Rev. B93~(15) (2016) 155106.
\newblock \href {http://arxiv.org/abs/1511.04918} {\path{arXiv:1511.04918}},
  \href {https://doi.org/10.1103/PhysRevB.93.155106}
  {\path{doi:10.1103/PhysRevB.93.155106}}.

\bibitem{Krieg:2018pqh}
S.~Krieg, T.~Luu, J.~Ostmeyer, P.~Papaphilippou, C.~Urbach, {Accelerating
  Hybrid Monte Carlo simulations of the Hubbard model on the hexagonal
  lattice}, Comput. Phys. Commun. 236 (2019) 15--25.
\newblock \href {http://arxiv.org/abs/1804.07195} {\path{arXiv:1804.07195}},
  \href {https://doi.org/10.1016/j.cpc.2018.10.008}
  {\path{doi:10.1016/j.cpc.2018.10.008}}.

\bibitem{Wynen:2018ryx}
J.-L. Wynen, E.~Berkowitz, C.~Körber, T.~A. Lähde, T.~Luu, {Avoiding
  Ergodicity Problems in Lattice Discretizations of the Hubbard Model}, Phys.
  Rev. B100~(7) (2019) 075141.
\newblock \href {http://arxiv.org/abs/1812.09268} {\path{arXiv:1812.09268}},
  \href {https://doi.org/10.1103/PhysRevB.100.075141}
  {\path{doi:10.1103/PhysRevB.100.075141}}.

\bibitem{pakman2015auxiliaryvariable}
A.~Pakman, L.~Paninski, {Auxiliary-variable Exact Hamiltonian Monte Carlo
  Samplers for Binary Distributions} (2015).
\newblock \href {http://arxiv.org/abs/1311.2166} {\path{arXiv:1311.2166}}.

\bibitem{similar_work}
Y.~Zhang, Z.~Ghahramani, A.~J. Storkey, C.~Sutton,
  \href{https://proceedings.neurips.cc/paper/2012/file/c913303f392ffc643f7240b180602652-Paper.pdf}{{Continuous
  Relaxations for Discrete Hamiltonian Monte Carlo}}, in: F.~Pereira, C.~J.~C.
  Burges, L.~Bottou, K.~Q. Weinberger (Eds.), Advances in Neural Information
  Processing Systems, Vol.~25, Curran Associates, Inc., 2012.
\newline\urlprefix\url{https://proceedings.neurips.cc/paper/2012/file/c913303f392ffc643f7240b180602652-Paper.pdf}

\bibitem{templates}
R.~Barrett, M.~Berry, T.~F. Chan, J.~Demmel, J.~Donato, J.~Dongarra,
  V.~Eijkhout, R.~Pozo, C.~Romine, H.~V. der Vorst, {Templates for the Solution
  of Linear Systems: Building Blocks for Iterative Methods}, SIAM,
  Philadelphia, PA, 1993.

\bibitem{PhysRevE.65.056706}
I.~P. Omelyan, I.~M. Mryglod, R.~Folk,
  \href{https://link.aps.org/doi/10.1103/PhysRevE.65.056706}{Optimized
  verlet-like algorithms for molecular dynamics simulations}, Phys. Rev. E 65
  (2002) 056706.
\newblock \href {https://doi.org/10.1103/PhysRevE.65.056706}
  {\path{doi:10.1103/PhysRevE.65.056706}}.
\newline\urlprefix\url{https://link.aps.org/doi/10.1103/PhysRevE.65.056706}

\bibitem{OMELYAN2003272}
I.~Omelyan, I.~Mryglod, R.~Folk,
  \href{http://www.sciencedirect.com/science/article/pii/S0010465502007543}{Symplectic
  analytically integrable decomposition algorithms: classification, derivation,
  and application to molecular dynamics, quantum and celestial mechanics
  simulations}, Computer Physics Communications 151~(3) (2003) 272 -- 314.
\newblock \href {https://doi.org/https://doi.org/10.1016/S0010-4655(02)00754-3}
  {\path{doi:https://doi.org/10.1016/S0010-4655(02)00754-3}}.
\newline\urlprefix\url{http://www.sciencedirect.com/science/article/pii/S0010465502007543}

\bibitem{ising_2d_exact}
A.~E. Ferdinand, M.~E. Fisher,
  \href{https://link.aps.org/doi/10.1103/PhysRev.185.832}{{Bounded and
  Inhomogeneous Ising Models. I. Specific-Heat Anomaly of a Finite Lattice}},
  Phys. Rev. 185 (1969) 832--846.
\newblock \href {https://doi.org/10.1103/PhysRev.185.832}
  {\path{doi:10.1103/PhysRev.185.832}}.
\newline\urlprefix\url{https://link.aps.org/doi/10.1103/PhysRev.185.832}

\bibitem{negele1988quantum}
J.~Negele, H.~Orland,
  \href{https://books.google.de/books?id=EV8sAAAAYAAJ}{Quantum many-particle
  systems}, Frontiers in physics, Addison-Wesley Pub. Co., 1988.
\newline\urlprefix\url{https://books.google.de/books?id=EV8sAAAAYAAJ}

\bibitem{monte_carlo_errors}
U.~{Wolff}, {Alpha Collaboration}, {Monte Carlo errors with less errors},
  Computer Physics Communications 156 (2004) 143--153.
\newblock \href {http://arxiv.org/abs/hep-lat/0306017}
  {\path{arXiv:hep-lat/0306017}}, \href
  {https://doi.org/10.1016/S0010-4655(03)00467-3}
  {\path{doi:10.1016/S0010-4655(03)00467-3}}.

\bibitem{Pelissetto:2000ek}
A.~Pelissetto, E.~Vicari, {Critical phenomena and renormalization group
  theory}, Phys. Rept. 368 (2002) 549--727.
\newblock \href {http://arxiv.org/abs/cond-mat/0012164}
  {\path{arXiv:cond-mat/0012164}}, \href
  {https://doi.org/10.1016/S0370-1573(02)00219-3}
  {\path{doi:10.1016/S0370-1573(02)00219-3}}.

\bibitem{Aarts:2010gr}
G.~Aarts, K.~Splittorff, {Degenerate distributions in complex Langevin
  dynamics: one-dimensional QCD at finite chemical potential}, JHEP 08 (2010)
  017.
\newblock \href {http://arxiv.org/abs/1006.0332} {\path{arXiv:1006.0332}},
  \href {https://doi.org/10.1007/JHEP08(2010)017}
  {\path{doi:10.1007/JHEP08(2010)017}}.

\bibitem{Fodor:2015doa}
Z.~Fodor, S.~D. Katz, D.~Sexty, C.~T{\"o}r{\"o}k, {Complex Langevin dynamics
  for dynamical QCD at nonzero chemical potential: a comparison with
  multi-parameter reweighting} (2015).
\newblock \href {http://arxiv.org/abs/1508.05260} {\path{arXiv:1508.05260}}.

\bibitem{Cristoforetti:2012su}
M.~Cristoforetti, F.~Di~Renzo, L.~Scorzato, {New approach to the sign problem
  in quantum field theories: High density QCD on a Lefschetz thimble}, Phys.
  Rev. D86 (2012) 074506.
\newblock \href {http://arxiv.org/abs/1205.3996} {\path{arXiv:1205.3996}},
  \href {https://doi.org/10.1103/PhysRevD.86.074506}
  {\path{doi:10.1103/PhysRevD.86.074506}}.

\bibitem{Fujii:2013sra}
H.~Fujii, D.~Honda, M.~Kato, Y.~Kikukawa, S.~Komatsu, T.~Sano, {Hybrid Monte
  Carlo on Lefschetz thimbles - A study of the residual sign problem}, JHEP 10
  (2013) 147.
\newblock \href {http://arxiv.org/abs/1309.4371} {\path{arXiv:1309.4371}},
  \href {https://doi.org/10.1007/JHEP10(2013)147}
  {\path{doi:10.1007/JHEP10(2013)147}}.

\bibitem{Alexandru:2017oyw}
A.~Alexandru, G.~Basar, P.~F. Bedaque, N.~C. Warrington, {Tempered transitions
  between thimbles}, Phys. Rev. D96~(3) (2017) 034513.
\newblock \href {http://arxiv.org/abs/1703.02414} {\path{arXiv:1703.02414}},
  \href {https://doi.org/10.1103/PhysRevD.96.034513}
  {\path{doi:10.1103/PhysRevD.96.034513}}.

\bibitem{Alexandru:2017czx}
A.~Alexandru, P.~F. Bedaque, H.~Lamm, S.~Lawrence, {Deep Learning Beyond
  Lefschetz Thimbles}, Phys. Rev. D96~(9) (2017) 094505.
\newblock \href {http://arxiv.org/abs/1709.01971} {\path{arXiv:1709.01971}},
  \href {https://doi.org/10.1103/PhysRevD.96.094505}
  {\path{doi:10.1103/PhysRevD.96.094505}}.

\bibitem{Mori:2017pne}
Y.~Mori, K.~Kashiwa, A.~Ohnishi, {Toward solving the sign problem with path
  optimization method}, Phys. Rev. D96~(11) (2017) 111501.
\newblock \href {http://arxiv.org/abs/1705.05605} {\path{arXiv:1705.05605}},
  \href {https://doi.org/10.1103/PhysRevD.96.111501}
  {\path{doi:10.1103/PhysRevD.96.111501}}.

\bibitem{Kashiwa:2018vxr}
K.~Kashiwa, Y.~Mori, A.~Ohnishi, {Control the model sign problem via path
  optimization method: Monte-Carlo approach to QCD effective model with
  Polyakov loop} (2018).
\newblock \href {http://arxiv.org/abs/1805.08940} {\path{arXiv:1805.08940}}.

\end{thebibliography}

\end{document}